\documentclass[apjl]{emulateapj}

\usepackage{natbib}
\usepackage[usenames]{color}

\newcommand{\bd}{\begin{displaymath}}
\newcommand{\ed}{\end{displaymath}}
\newcommand{\be}{\begin{equation}}
\newcommand{\ee}{\end{equation}}
\newcommand{\beaa}{\begin{eqnarray*}}
\newcommand{\eeaa}{\end{eqnarray*}}
\newcommand{\bea}{\begin{eqnarray}}
\newcommand{\eea}{\end{eqnarray}}

\newcommand\rxj{RXJ1131$-$1231}
\newcommand\rxjshort{RXJ1131}
\newcommand\blens{B1608$+$656}

\def\hst{\textit{HST}}

\def\Ok{\Omega_{\rm k}}
\def\Ode{\Omega_{\rm de}}
\def\Om{\Omega_{\rm m}}

\def\slope{\gamma'}
\def\thEin{\theta_{\rm E}}
\def\kext{\kappa_{\rm ext}}
\def\gext{\gamma_{\rm ext}}
\def\tdist{D_{\Delta t}}
\def\tdistmod{D_{\Delta t}^{\rm model}}

\def\rs{r_{\rm s}}

\def\im{\theta}

\def\kms {\rm km\,s^{-1}}
\def\kmsMpc {\rm km\,s^{-1}\,Mpc^{-1}}

\def\Dtfitlam{1388.8}
\def\Dtfitmu{6.4682}
\def\Dtfitsig{0.20560}

\newcommand{\sref}[1]{Section~\ref{#1}}
\newcommand{\fref}[1]{Figure~\ref{#1}}

\newcommand{\eref}[1]{Equation~(\ref{#1})}

\begin{document}

\title{Cosmology from gravitational lens time delays and Planck data}

\author{S.~H.~Suyu\altaffilmark{1},
T.~Treu\altaffilmark{2,*},
S.~Hilbert\altaffilmark{3},
A.~Sonnenfeld\altaffilmark{2},
M.~W.~Auger\altaffilmark{4},
R.~D.~Blandford\altaffilmark{5},
T.~Collett\altaffilmark{4},
F.~Courbin\altaffilmark{6},
C.~D.~Fassnacht\altaffilmark{7},
L.~V.~E.~Koopmans\altaffilmark{8},
P.~J.~Marshall\altaffilmark{9,5},
G.~Meylan\altaffilmark{6},
C.~Spiniello\altaffilmark{8,3},
M.~Tewes\altaffilmark{6,10}}

\altaffiltext{1}{Institute of Astronomy and Astrophysics, Academia Sinica, P.O.~Box 23-141, Taipei 10617, Taiwan}
\altaffiltext{2}{Department of Physics, University of California, Santa Barbara, CA 93106, USA} 
\altaffiltext{3}{Max-Planck-Institut f{\"u}r Astrophysik, Karl-Schwarzschild-Str.~1, 85748 Garching, Germany}
\altaffiltext{4}{Institute of Astronomy, University of Cambridge, Madingley Rd, Cambridge, CB3 0HA, UK}
\altaffiltext{5}{Kavli Institute for Particle Astrophysics and Cosmology, Stanford University, 452 Lomita Mall, Stanford, CA 94035, USA} 
\altaffiltext{6}{Laboratoire d'Astrophysique, Ecole Polytechnique F{\'e}d{\'e}rale de Lausanne (EPFL), Observatoire de Sauverny, CH-1290 Versoix, Switzerland}
\altaffiltext{7}{Department of Physics, University of California, Davis, CA 95616, USA}
\altaffiltext{8}{Kapteyn Astronomical Institute, University of Groningen, P.O.Box 800, 9700 AV Groningen, The Netherlands}
\altaffiltext{9}{Department of Physics, University of Oxford, Keble Road, Oxford, OX1 3RH, UK}
\altaffiltext{10}{Argelander-Institut f\"ur Astronomie, Auf dem H\"ugel 71, 53121 Bonn, Germany}
\altaffiltext{*}{Packard Research Fellow}

\email{suyu@asiaa.sinica.edu.tw}

\shorttitle{Gravitational lens time delays and Planck}
\shortauthors{Suyu et al.}

%-------------------------------------------------------------------------------

\begin{abstract}

Under the assumption of a flat $\Lambda$CDM cosmology, recent data from the
Planck satellite point toward a Hubble constant that is in tension with that
measured by gravitational lens time delays and by the local distance ladder.
Prosaically, this difference could arise from unknown systematic uncertainties
in some of the measurements. More interestingly -- if systematics were ruled
out -- resolving the tension would require a departure from the flat
$\Lambda$CDM cosmology, 
introducing for example a modest amount of spatial curvature, or a non-trivial
dark energy equation of state.  
To begin to address these issues, we present here an analysis of the
gravitational lens \rxj\ that is improved in one particular regard: 
we examine the issue of systematic error introduced by an assumed lens
model density profile.  We use more flexible
gravitational lens models with baryonic and dark matter components, and
find that the exquisite \textit{Hubble Space Telescope} image with
thousands of intensity pixels in the Einstein ring and the stellar
velocity dispersion of the lens contain sufficient
information to constrain these more flexible models. The total
uncertainty on the time-delay distance is $6.6\%$ for a
single system.
We proceed to combine our improved time-delay
distance measurements with the WMAP9 and Planck posteriors.  
In an open
$\Lambda$CDM model, the data for \rxj\ in combination with Planck
favor a flat universe with $\Ok=0.00^{+0.01}_{-0.02}$ (68\% CI).  In a flat
$w$CDM model, the combination of \rxj\ 
and Planck yields $w=-1.52^{+0.19}_{-0.20}$ (68\% CI). 

\end{abstract}
 

%-------------------------------------------------------------------------------

\section{Introduction} 
\label{sec:intro}

The last few years have been hailed as the era of precision
cosmology. Many different methods now point to the so-called
concordance cosmology, characterized by a virtually flat geometry in a
universe dominated by dark matter and dark energy
\citep[e.g.,][]{HinshawEtal12,Planck2013P16}. 
With precision on many parameters now reaching the few percent level,
it is extremely valuable to compare and contrast different probes. A
comparison between independent probes
is the cleanest way to test the accuracy of the
measurements. Furthermore, certified tension between independent
probes' measurements 
would require the falsification of the simplest models and potentially
the discovery of new physics \citep{SuyuEtal12b}.

A classic example is the interpretation of the cosmic microwave background
(CMB) data. The power spectrum of the CMB anisotropies delivers an
enormous amount of information about the high-redshift universe, but 
it is not directly sensitive 
to lower-redshift phenomena. 
Thus,
inferring $w$ or the Hubble constant ($H_0$) from the CMB data typically
requires strong assumptions about the cosmological model
(e.g.~flatness) or the combination with lower redshift probes.  This
is well exemplified by the Planck analysis 
\citep[Paper XVI;][]{Planck2013P16}. Assuming $\Ok=0$ and $w=-1$, $H_0=67.3\pm1.2\,\kmsMpc$, in
tension with that measured by various lower-redshift methods
\citep[e.g.,][]{RiessEtal11,FreedmanEtal12, ChavezEtal12,
  SuyuEtal13}.  
If confirmed, this
tension would imply that the simplest flat $\Lambda$CDM is
falsified. Given the high stakes, it is crucial to
re-examine the uncertainties of each method, eliminating unaccounted
for systematics.

The aim of this paper is two-fold. Firstly, we present a re-analysis
of the gravitational lens system \rxj\ (\fref{fig:rxj}) discovered by
\citet{SluseEtal03}.  Following 
the work of \citet[][hereafter 
SS13]{SchneiderSluse13} we consider composite mass models for the main
deflector galaxy (\sref{sec:lensmod}). The composite models consist
of stellar and dark matter components, and are thus more realistic and
flexible than the power-law models considered in 
our original 
analysis \citep[][hereafter SU13]{SuyuEtal13}. We show that even with this broader class of lens
models, our deep \textit{Hubble Space Telescope} (\hst) images of the Einstein
ring together with the stellar velocity dispersion measurement of the
lens allow us to constrain the time-delay distance ($\tdist
\propto H_0^{-1}$), a combination of angular diameter distances.

Secondly, having shown that uncertainties in the mass model are not
significantly larger than our previous estimate, we proceed in
\sref{sec:cosmo} to combine our $\tdist$ measurement with
the recent CMB results from the Wilkinson Microwave
Anisotropy Probe 9-year data \citep[WMAP9;][]{HinshawEtal12} and from
Planck \citep{Planck2013P16}.  
We conclude in \sref{sec:conclude}.

Throughout this paper, each quoted parameter estimate is the median of
the marginalized posterior probability density
function (PDF), with the uncertainties showing the $16^{\rm th}$ and
$84^{\rm th}$ percentiles (i.e., the 68\% credible interval (CI)).

%-------------------------------------------------------------------------------
\section{Lens mass models: power of spatially extended Einstein rings}
\label{sec:lensmod}

\begin{figure}
  \centering
  \includegraphics[width=0.45\textwidth, clip]{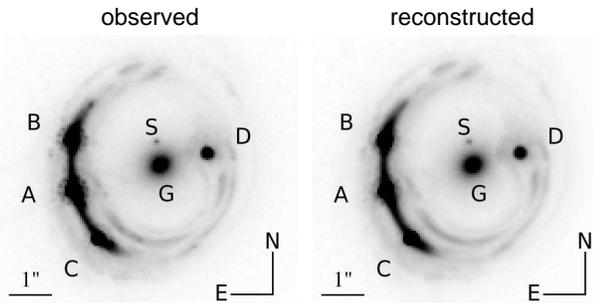}
  \caption{\label{fig:rxj} \hst\ ACS
    image of \rxj\ in F814W filter.  The background AGN is lensed into
    four images (A, B, C and D) by the primary lens galaxy G and its
    satellite S.  Left: observed image.  Right: reconstructed image
    based on the most probable composite model in
    \sref{sec:lensmod:composite}.}
\end{figure}

SU13 modeled the lens galaxy in \rxj\ with a power-law mass distribution that
was motivated by several studies, including the X-ray observations of
galaxies \citep{HumphreyBuote10} and the Sloan Lens ACS survey
\citep[e.g.,][]{KoopmansEtal06, GavazziEtal07,KoopmansEtal09, AugerEtal10,
BarnabeEtal11}, which found that galaxies are well described by 
power-law mass distributions in regions covered by the data.  Furthermore, the pixelated lens
potential corrections applied by \citet{SuyuEtal09} to the
gravitational lens \blens\ was within $\sim$$2\%$ from 
a power law, validating the use of a simple
power-law model.  Here, we assess further the dependence of 
$\tdist$ on the form of the mass model by 
employing two other forms that were 
considered by SS13: a cored power-law mass distribution, and a
composite model of dark matter and baryons.  In each case,
we use the time delays from \citet{TewesEtal13b}\footnote{based on
  monitorings of COSMOGRAIL \citep[COSmological MOnitoring of
  GRAvItational Lenses; e.g.,][]{CourbinEtal11,
    TewesEtal13} and \citet{KochanekEtal06} teams.}
and the \hst\ image (\fref{fig:rxj}; SU13) 
to constrain the lens model.  The expressions 
for the likelihoods of the data are given in Section 6.2 of SU13.

\subsection{Cored power-law model}
\label{sec:lensmod:coredpl}

The dimensionless surface mass density (convergence) of a cored
elliptical power-law profile is given by
\be
\label{eq:coredPLkappa}
\kappa_{\rm cpl}(\im_1,\im_2) =
\frac{3-\slope}{2}\left(\frac{\thEin}{\sqrt{q \im_1^2+\im_2^2/q + \im_{\rm c}^2}} \right)^{\slope-1},
\ee
where $(\im_1,\im_2)$ are coordinates on the lens/image plane, 
$\slope$ is the radial power-law slope (with $\slope=2$
corresponding to isothermal), $\thEin$ is the Einstein radius, $q$ is
the axis ratio, and $\im_{\rm
c}$ is the core radius.  This is identical to the lens mass
distribution in SU13 except for the non-zero $\im_{\rm c}$ here.

\fref{fig:rxj} shows a primary lens galaxy G and a
satellite lens galaxy S that are surrounded by the Einstein ring of
the lensed source.  Following SU13 in the modeling procedure, we
remodel the Advanced Camera for Surveys (ACS) image using the cored 
power-law profile for G.  For simplicity, we fix the mass
distribution of S to that of the most probable model in
SU13 since the satellite impacts the $\tdistmod$
measurement at the $<$$1\%$ level.  We also include an external shear
contribution with strength $\gamma_{\rm ext}$ and position angle
$\phi_{\rm ext}$.  We use a grid of $50\times50$ 
intensity pixels on the source plane to model the spatially extended quasar host galaxy.
These source pixels map to an annular region on the image
  plane containing the arcs
  that are visible in \fref{fig:rxj}.  
We sample the lens parameters and $\tdistmod$ using
the same Markov chain Monte Carlo (MCMC) methods as were used in SU13.
The lensing data constrain the maximum $\im_{\rm c}$ to be $0.005''$ (95\%
CI), and $\im_{\rm c}=0$ is compatible with the data.  
The marginalized values of the lens
parameters, $\tdistmod$, and the cosmological results are the same as those
presented in SU13 within two significant digits.

\subsection{Composite mass model}
\label{sec:lensmod:composite}

In the composite model, we treat baryons and dark matter
individually.  We model the baryonic mass distribution of the lens
galaxy G as its observed light profile  normalized by a constant $M/L$.
The difference of two isothermal profiles mimics a Sersic 
profile \citep{DuttonEtal11} and provides efficient computation of
lensing quantities:
\bea
\label{eq:mimicSersic}
L(\im_1,\im_2) & =& \frac{L_0}{(1+q_{\rm
    L})} {\Big [}\frac{1}{\sqrt{\im_1^2+\im_2^2/{q_{\rm L}}^2 + 4w_{\rm
      c}^2/(1+q_{\rm L})^2}}
  \nonumber \\
&& - \frac{1}{\sqrt{\im_1^2+\im_2^2/{q_{\rm L}}^2 + 4w_{\rm t}^2/(1+q_{\rm L})^2}} { \Big ]},
\eea
where $q_{\rm L}$ is the axis ratio, and $w_{\rm c}$ and $w_{\rm t}$ are
profile parameters with $w_{\rm t} > w_{\rm c}$.  We use two sets of
the above profile with common centroid and position angle to fit the
light distribution of G in the ACS image  
since a single one is inadequate (\citealt{ClaeskensEtal06}; SU13).  
The optimized structural parameters are
  $(q_{\rm L1}, w_{\rm c1}, w_{\rm t1})=(0.88, 2.0, 2.5)$ and $(q_{\rm
    L2}, w_{\rm c2}, w_{\rm t2})=(0.85,0.06,0.67)$, and are held fixed
  since the uncertainties on these parameters
  ($<2\%$) are negligible in terms of their effect on
  $\tdistmod$ (SU13).
For the dark
matter halo, we adopt the standard NFW profile
\citep{NavarroEtal96} whose three-dimensional density is 
\be
\label{eq:nfw}
\rho(r) = \frac{\rho_0}{(r/\rs) (1+r/\rs)^2},
\ee
where $\rho_0$ is a normalization and $\rs$ is the scale radius.  We
follow \citet{GolseKneib02} for obtaining the deflection angles and
lens potential of an elliptical NFW profile in
projection.\footnote{\citet{GolseKneib02} introduced the ellipticity
into the lens potential, and \citet{SandEtal08} showed that this
yields valid elliptical surface mass density when $q_{\rm
h}\gtrsim0.8$.}  For the satellite, we model its mass distribution as a singular isothermal sphere centered on its light distribution.  As in the previous cases, we allow for an external
shear contribution.

We have 11 parameters in modeling the ACS image and time
delays: a global $M/L$ of the baryons, the NFW parameters (centroid
$(\im_{\rm 1h},\im_{\rm 2h})$, axis ratio $q_{\rm h}$, position angle
$\phi_{\rm h}$, normalization $\kappa_{\rm 0,h}$, scale radius $\rs$), satellite Einstein radius $\im_{\rm E,S}$, 
external shear $\gamma_{\rm ext}$ and $\phi_{\rm ext}$, and the
modeled time-delay distance $\tdistmod$.  
We allow the centroid of the NFW halo to vary from the lens galaxy
  G with Gaussian uncertainties of $\pm0.01''$.  We adopt a Gaussian prior on $\rs$ of $18.6''\pm2.6''$
based on the weak lensing analysis of the SLACS lenses
\citep{GavazziEtal07} that have similar velocity dispersions as that
of \rxj.  For the other parameters, we impose uniform
priors. 

\begin{figure}
  \centering
 \includegraphics[width=0.35\textwidth, clip, angle=270]{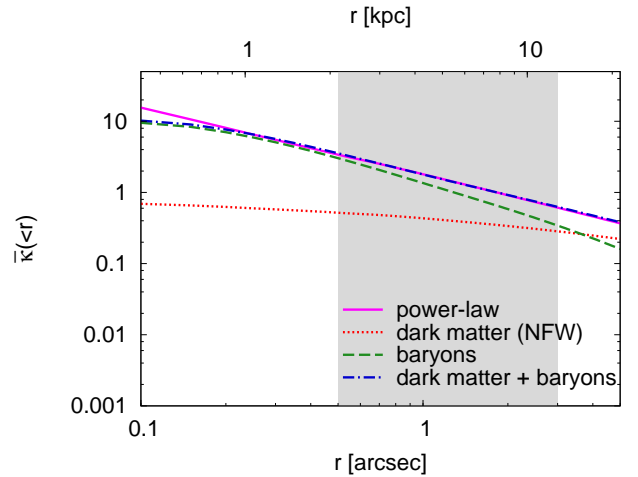}
  \caption{\label{fig:meankvsr} The circularly averaged convergence as
    a function of radius for the most probable models.  The power-law
    model (solid) is from SU13.  The composite
    model (dot-dashed) consists of a baryonic mass distribution based
    on the light profile (dashed), and a dark matter distribution
    based on an NFW profile (dotted).  The convergence includes
    the contribution from the satellite galaxy.  In the region covered by the Einstein  
    ring, between $\sim0.5''$ and $\sim3''$, the slope of the
    composite model is nearly identical to that of the single
    power law.  The spatially extended Einstein ring covering thousands
    of intensity pixels provide strong constraints on the local
    lens mass profile.}
\end{figure}

We sample the 11 parameters using MCMC for a series of source
intensity grids: $50\times50$, $52\times52$, $54\times54$,
$56\times56$, $58\times58$, $60\times60$ and $64\times64$. As in SU13,
the effects of the source grid resolution dominate the uncertainty on the
lens parameters.  We conservatively combine the results of the
different source resolutions by weighting each equally and
approximating the combined PDF with a multivariate Gaussian.  In the
right-hand panel of \fref{fig:rxj}, we show the reconstructed \hst\
image based on our most probable composite model with $64\times64$
source pixels, which reproduces the global features of the observed
image.  In \fref{fig:meankvsr}, we show the circularly averaged
convergence of the same model.  
Within the shaded region spanned
by the spatially extended arcs, the combination of
the baryons (dashed) and the dark matter (dotted) in the composite
model yields a nearly perfect power-law profile (dot-dashed). For comparison, the
power-law model from SU13 is also plotted in solid.  Therefore, the
spatially extended arcs and the time delays provide strong constraints
on the local profile of the lens mass distribution.

The composite model requires an external shear strength of
$\gamma_{\rm ext}=0.075\pm0.005$ at an angle of $80\pm3\degr$
  that is overall consistent with the 
distribution of external mass concentrations (see Figure 5 of SU13). 

%-------------------------------------------------------------------------------

\section{Impact on time-delay distance}
\label{sec:tdist}

\begin{figure}
  \centering
  \includegraphics[width=0.40\textwidth, clip]{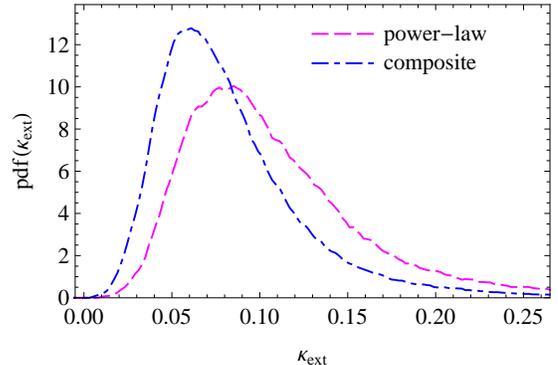}
  \caption{\label{fig:kextpdf} The PDF for the external convergence
    $\kext$ from combining ray tracing through the Millennium
    Simulation with (1) the galaxy count around the lens system that
    is 1.4 times the average number of galaxy counts, and (2) the
    modeled external shear.  The composite model with $\gext =
    0.075\pm0.005$ is in dot-dashed, and the power-law model with
    $\gext=0.089\pm0.006$ (SU13) is in dashed.}
\end{figure}

The cosmological $\tdist$ to the lens is affected by the external
mass distributions along the line of sight and is related to 
$\tdistmod$ by 
\be
\label{eq:tdist}
\tdist = \frac{\tdistmod}{1-\kext},
\ee
where $\kext$ characterizes the external convergence associated with
these mass structures.  Following SU13, we construct the PDF of $\kext$ by
ray tracing through the 
Millennium Simulation \citep{SpringelEtal05, HilbertEtal09}, selecting
lines of sight in the simulations with a galaxy count around the lens
system that is 1.4 times the average (measured by
\citealt{FassnachtEtal11}), and weighting by the external shear
value\footnote{Each selected line of sight
    from the simulation is weighted by the probability of its shear
    value given the measured value of $0.075\pm0.005$
    in Section \ref{sec:lensmod:composite}.}.   In
\fref{fig:kextpdf}, we show the PDFs of $\kext$ for the two models,
which differ by $\sim$$0.02$ due to the shear strengths. 

\begin{figure*}
  \centering
  \includegraphics[width=0.35\textwidth, clip]{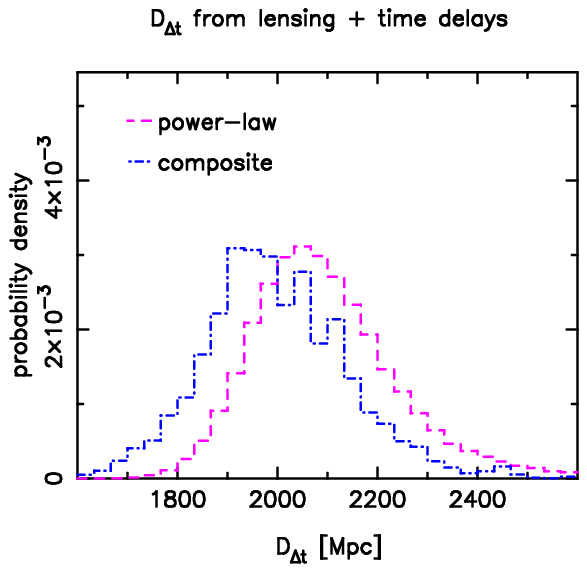}
  \includegraphics[width=0.35\textwidth, clip]{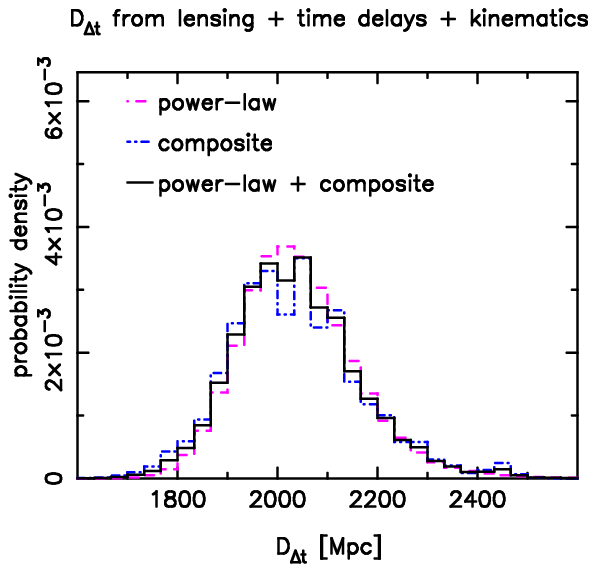}
  \caption{\label{fig:Ddt} The time-delay distance,
    $\tdist$, for the power-law model (dashed) and the
    composite model of baryons and dark matter (dot dashed) in the
    U$H_0$ cosmology.  The left-hand panel is based on only the
    lensing and time-delay data, whereas the right-hand panel
    includes the information from the lens velocity dispersion.  The
    stellar kinematic information on the lens galaxy help break lens
    model degeneracies, yielding very similar $\tdist$ distributions
    for the two lens models.  
    The combined PDF of $\tdist$ is shown in solid
    in the right-hand panel. } 
\end{figure*}

Combining the $\tdistmod$ with the
$\kext$ PDF for each of the two lens models, we obtain via
\eref{eq:tdist} the PDF for $\tdist$ based on the lensing and time
delay data (\fref{fig:Ddt} left-hand panel)\footnote{assuming a
  uniform prior on $H_0$.}.  The 
$\tdist$ PDFs for the two models are shifted with respect to each
other by $\sim 4\%$.

We use the measured lens velocity
dispersion of $323\pm20\,\kms$ within a rectangular aperture of
$0.81''\times0.7''$ (SU13) to further constrain the
lens models and help break the mass-sheet degeneracy
\citep[e.g.,][]{Koopmans04}.  The kinematic modeling of the power-law
model is described in SU13.  For the composite model, we follow
\citet{SonnenfeldEtal12} to model the velocity dispersion of the
baryonic and the dark matter distributions.  In the right-hand panel of
\fref{fig:Ddt}, we show the resulting $\tdist$ PDF by combining
lensing, time-delay, and lens-velocity-dispersion
measurements. The kinematic data help break lens model degeneracies and
provide robust $\tdist$ measurements that are less sensitive to lens
model assumptions.  We conservatively assign equal priors to the
two models (power-law and composite), on the grounds that we have no reason to
believe one parametrization over another {\it a priori}. The combined
$\tdist$ distribution is shown in solid lines. 

\smallskip

A fitting formula for the PDF of $\tdist$, which
can be used to combine with any other independent data set, is
\bea
\label{eq:DtLogNorm}
\lefteqn{P(\tdist|H_0,\Ode,w,\Om) \simeq }\nonumber\\
 && \frac{1}{\sqrt{2\pi} (x-\lambda_{\rm D}) \sigma_{\rm D}} 
\exp{\left[-\frac{(\log(x-\lambda_{\rm D}) - \mu_{\rm D})^2}{2\sigma_{\rm D}^2}\right]},
\eea
where $x=\tdist/(1\, {\rm Mpc})$, $\lambda_{\rm D} = \Dtfitlam$, $\mu_{\rm
  D}=\Dtfitmu$ and $\sigma_{\rm D} = \Dtfitsig$.  
Our inference of $\tdist$ is accurate to $\sim 6.6\%$.

%-------------------------------------------------------------------------------

\section{Cosmology with time-delay lenses and the CMB}
\label{sec:cosmo}

The time-delay distance allows us to infer cosmological parameters.
We consider five background cosmological models with four of them
based on the recent results from WMAP9 \citep{HinshawEtal12} and
Planck \citep{Planck2013P16}: (1) uniform $H_0$ (U$H_0$) in flat $\Lambda$CDM
with $\Omega_{\Lambda}=1-\Omega_{m}=0.73$, which is useful for
comparing to earlier lensing results, (2)
WMAP9 open $\Lambda$CDM, (3) WMAP9 $w$CDM, (4) Planck open
$\Lambda$CDM, and (5) Planck $w$CDM.  Compared to the flat
$\Lambda$CDM model, open $\Lambda$CDM allows for spatial curvature
$\Ok$, and $w$CDM allows for a time-independent $w$ that is not fixed
to $-1$.  We consider these more generic models in (2)-(5) given the current
tensions in $H_0$ measurements from Planck in flat $\Lambda$CDM
cosmology and from direct probes \citep{Planck2013P16}.  

\begin{figure}[t!]
  \vspace{0.8cm}
  \hspace{0.3cm}  WMAP9 \hspace{2.7cm}  Planck
  \centering
  \includegraphics[width=0.22\textwidth, clip]{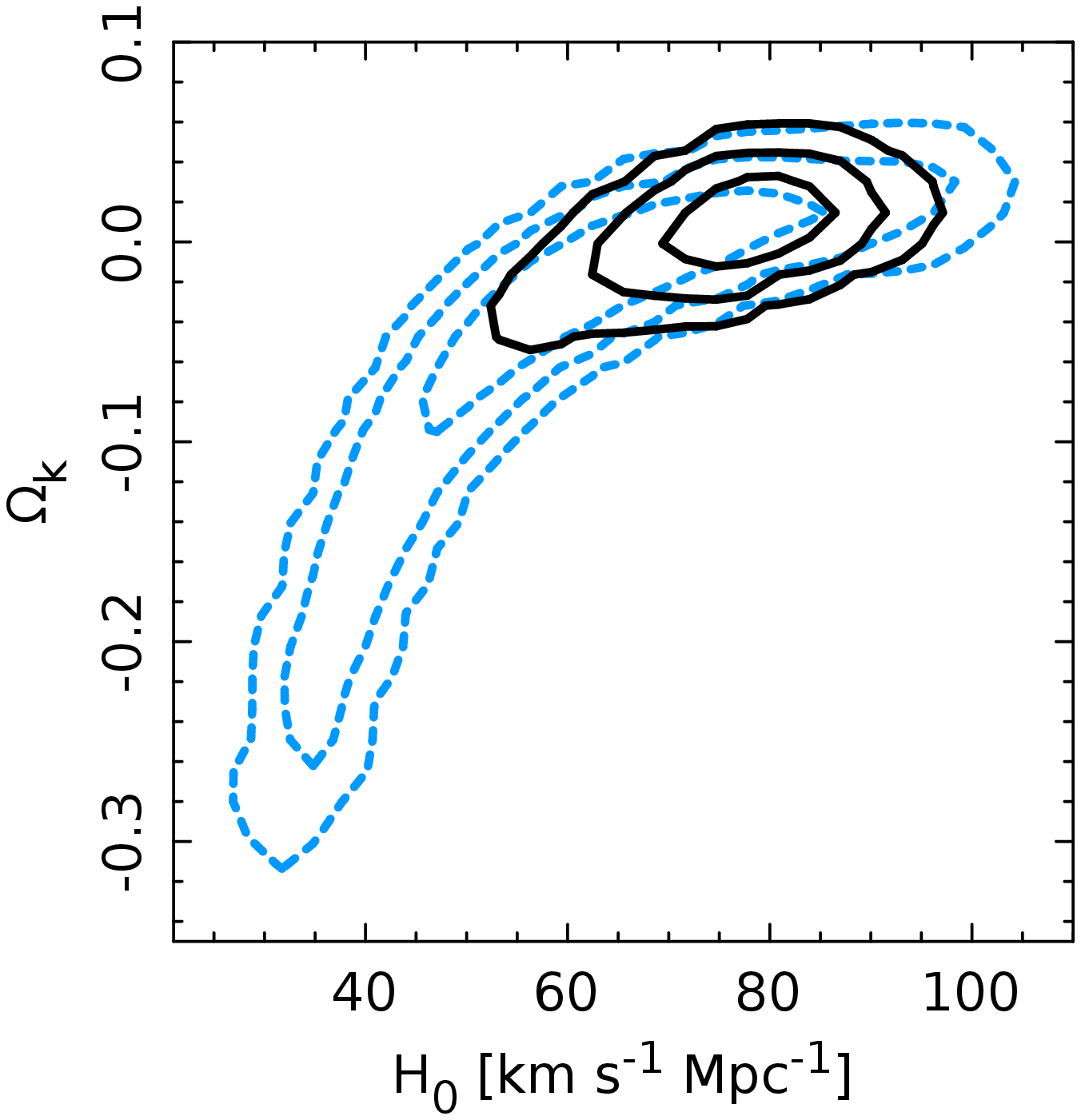}
  \includegraphics[width=0.22\textwidth, clip]{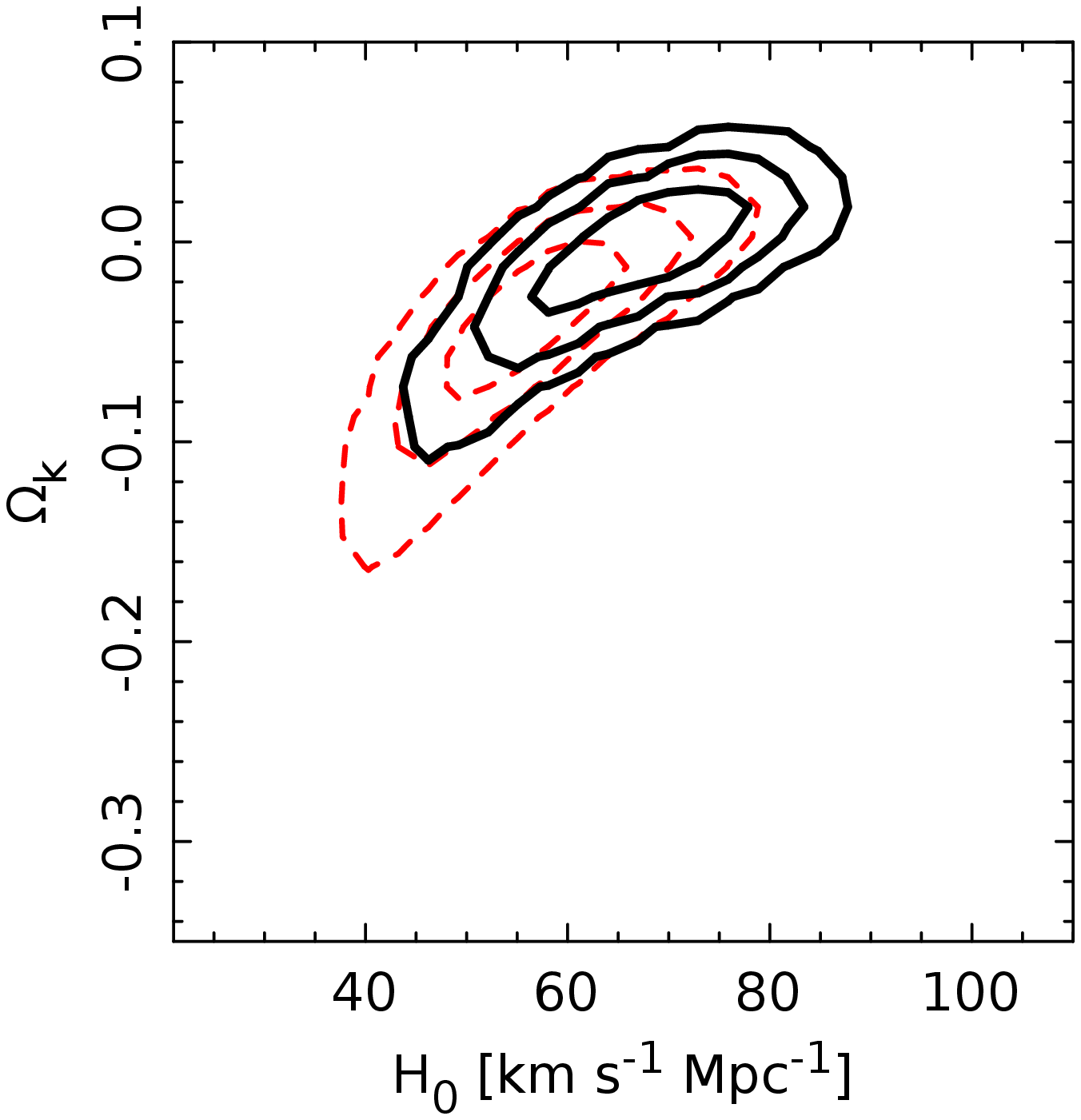}
  \includegraphics[width=0.22\textwidth, clip]{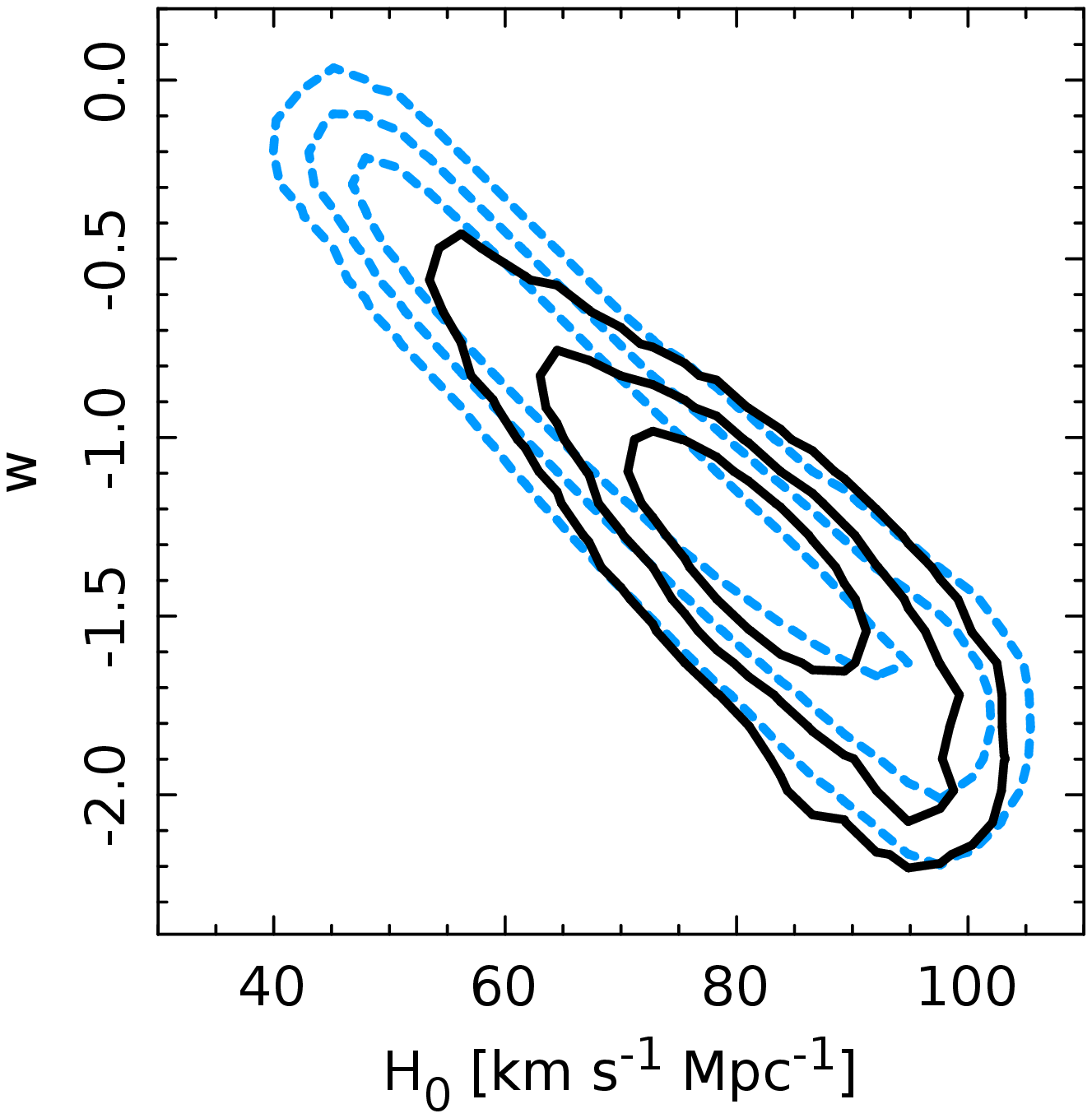}
  \includegraphics[width=0.22\textwidth, clip]{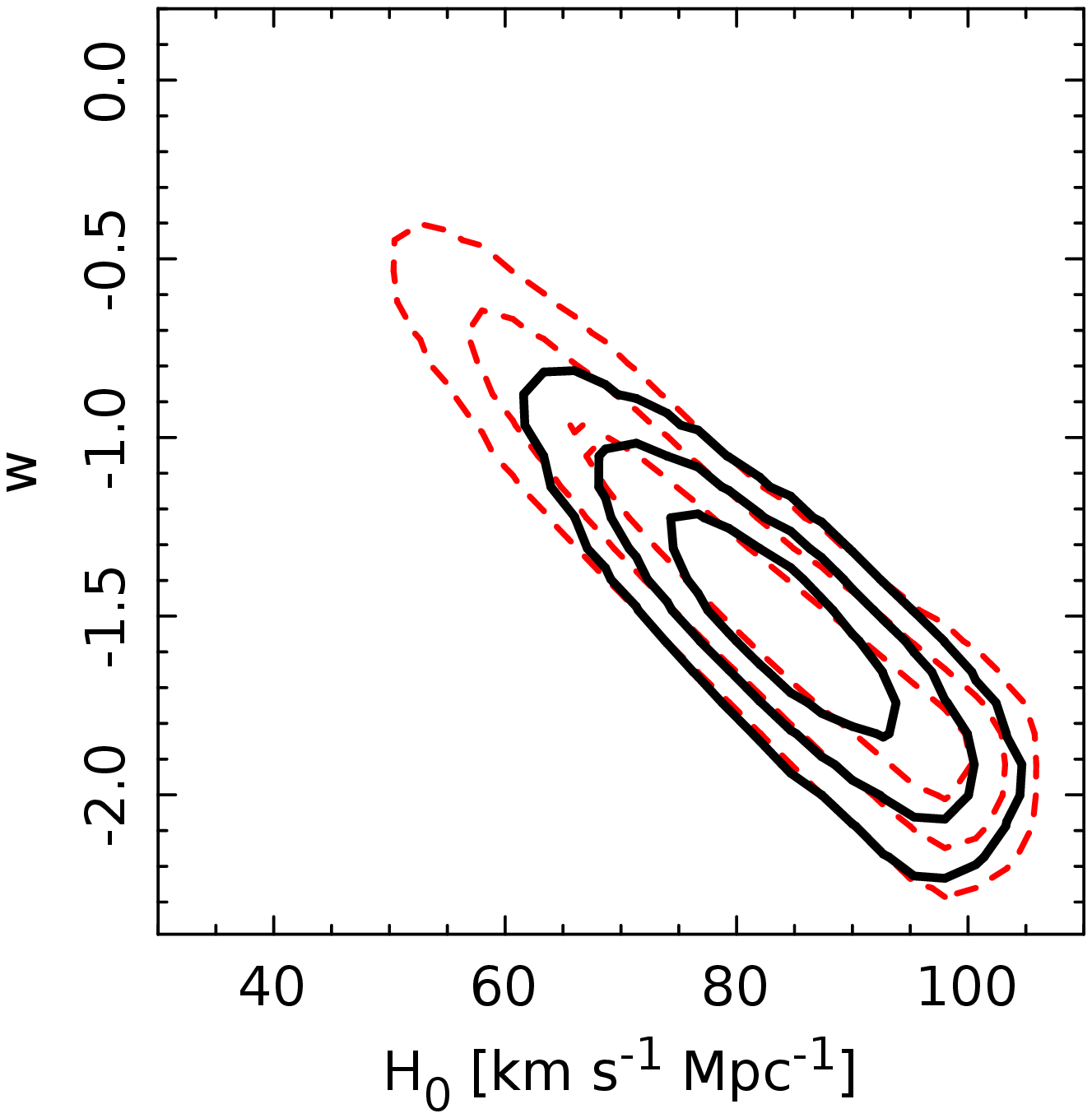}

  \caption{\label{fig:cosmo} Cosmological constraints assuming open
    $\Lambda$CDM (top) and $w$CDM (bottom).  Left/Right panels:
    WMAP9/Planck priors are shown in dashed, and the combination of
    \rxj\ with WMAP9/Planck is in solid.  \rxj, which primarily
    constrains $H_0$, helps break parameter degeneracies in the CMB to
    determine flatness and $w$.}
\end{figure}

For each of the five cosmological priors, we importance sample
  the parameters \{$H_0$, $\Omega_{\rm m}$, $\Omega_{\Lambda}$, $w$\}
  from the U$H_0$ prior or WMAP9/Planck MCMC chains\footnote{For the Planck
  chains, we use the ones from the Planck temperature power spectrum
  in combination with WMAP9 low-$l$ polarization data.}
(\citealt{LewisBridle02}; 
SU13) with the likelihood of the \rxj\ data from our improved analysis.  In
\fref{fig:cosmo}, we show the  
cosmological constraints from the combination of \rxj\ with WMAP9
(left-hand panels) or with Planck (right-hand panels) in solid
contours.  Compared to the WMAP9-only or Planck-only constraints
(dashed), the gravitational lens \rxj\ reduces the
parameter degeneracies in the CMB data.  

The constraint on $H_0$ in the U$H_0$ cosmology is
$H_0=80.0^{+4.5}_{-4.7}\,\kmsMpc$. 

The marginalized joint constraints in open $\Lambda$CDM are \\
$\left\{ \begin{array}{ll}
H_0 = 78.0^{+4.6}_{-5.1}\,\kmsMpc\\
\Ok = 0.011^{+0.006}_{-0.007} \\
\end{array} \right. $ (WMAP9+\rxjshort) \\
and \\
$\left\{ \begin{array}{ll}
H_0 = 67.3^{+6.1}_{-6.6}\,\kmsMpc \\
\Ok = 0.00^{+0.01}_{-0.02} \\
\end{array} \right. $ (Planck+\rxjshort). \\
The marginalized joint constraints in the flat $w$CDM model are \\
$\left\{ \begin{array}{ll}
H_0 = 81.4^{+6.2}_{-6.2} \,\kmsMpc\\
w = -1.33^{+0.20}_{-0.22} \\
\end{array} \right. $ (WMAP9+\rxjshort) \\
and \\
$\left\{ \begin{array}{ll}
H_0 = 84.2^{+6.4}_{-5.9} \,\kmsMpc\\
w = -1.52^{+0.19}_{-0.20} \\
\end{array} \right. $ (Planck+\rxjshort). \\
The difference in the above marginalized $H_0$ with Planck in the two
cosmologies is driven by the Planck data.   If we restrict $w\geq -1$ (the physical regime in most models), then we infer $w=-0.92^{+0.16}_{-0.05}$ with WMAP9+\rxjshort\ and $w=-0.94^{+0.06}_{-0.05}$ with Planck+\rxjshort.

%-------------------------------------------------------------------------------

\section{Discussions and Conclusion}
\label{sec:conclude}

The two dominant sources of uncertainty in determining $\tdist$
originate from (1) the radial profiles of the lens mass 
distribution in the region spanned by the images, and 
(2) weak lensing effects due to mass structures along the line of sight.
Recently, SS13 considered toy models of spherical lens mass
distributions with point-like sources and suggested that lens models
with different radial profiles can lead to $\tdist$ values that differ
by $\sim20\%$ while fitting to the point-like images.  
We have considered their two lens density profiles in an improved analysis
of \rxj, and have
demonstrated that the \textit{spatially extended Einstein ring of the
lensed source} and the availability of multiple time delays provide 
strong constraints on the \textit{local} profile of the lens mass distribution
(\fref{fig:meankvsr}).  
By incorporating the lens velocity dispersion measurement and estimates of the external convergence $\kext$, 
we break degeneracies in the lens model (\fref{fig:Ddt}), 
and show that the results are robust with respect to the chosen form of mass
profile at large radii (close to isothermal versus NFW).  Work is
underway to improve estimates of $\kext$ \citep[e.g.,][]{GreeneEtal13,
  CollettEtal13, McCullyEtal14}. 

By modeling the baryons separately from the dark matter halo, we
obtain a rest-frame $M/L_{\rm V}=7\pm3 \,M_{\sun}/L_{\rm V,\sun}$ 
for the baryonic
component, where the uncertainty stems mainly from the
extrapolation of the lens light profile at large radii.  
The dark
matter mass fraction ($f_{\rm DM}$) within the lens galaxy effective radius of
$1\farcs85$ is $\sim 35\%$.  
These values 
are typical of massive early-type galaxies 
\citep[e.g.,][]{AugerEtal10, BarnabeEtal11}.  As seen in
\fref{fig:meankvsr}, neither the dark matter nor the baryons is a
power law, but the combination of the two leads to a nearly perfect
power law locally.  This ``bulge-halo conspiracy'' has already been
noted in earlier studies \citep[e.g.,][]{TreuKoopmans04,
KoopmansEtal09, vandeVenEtal09} and is reproduced by some numerical
simulations including baryonic physics \citep{RemusEtal13}.

The centroids of the dark matter halo and the baryonic
component of the primary lens galaxy are offset by $\sim 0.1''$, while
their position 
angles agree within $6\degr$.  This suggests
that the surface mass density of the lens is more complex than a simple
elliptical distribution, which is not surprising given the presence of
the satellite galaxy.
Despite this, the inference of $\tdist$ is robust: the 
various lens model
assumptions lead to similar $\tdist$, within $\sim2\%$, given the
exquisite data set.  We give a fitting formula in \eref{eq:DtLogNorm}
for the PDF of the inferred $\tdist$ to \rxj\ that can be combined
with any independent probe of cosmology.

The inferred $H_0$ value from Planck in the flat $\Lambda$CDM model is
$67.3\pm1.2\,\kmsMpc$.  This is in tension with several direct $H_0$
probes, including the Cepheids distance ladder with
$H_0=73.8\pm2.4\,\kmsMpc$ \citep{RiessEtal11} or
$H_0=74.3\pm1.5{\rm\,(stat.)}\pm2.1{\rm\,(sys.)}\,\kmsMpc$
\citep{FreedmanEtal12}.  Our measurement of $H_0$ from \rxj\  is also
in tension with 
the Planck value under the flat $\Lambda$CDM assumption.  We emphasize
that the $H_0$ measurements from the CMB are highly model dependent
and can change markedly when one relaxes from spatial
flatness or $\Lambda$ (\fref{fig:cosmo}).  The
currently perceived tension could be due to unknown systematic
uncertainties or an indication of new physics such as the dark energy
component not having $w=-1$.  It is now
crucial to pin down the uncertainties of each approach and
employ multiple independent probes to rule out unknown systematics.

Gravitational lens time delays provide an independent one-step method
to determine cosmological distances.  With extensive data sets on
\rxj, we measure its $\tdist$ to a precision of $6.6\%$. 
We will soon have three more time-delay lenses with similar data quality as that of \rxj\ and
\blens\ to reduce
our statistical uncertainties on cosmological parameters, and more
importantly, to test for the presence of residual systematics in our
approach. 
By understanding and eliminating our systematic
uncertainties, the statistical power of the hundreds of time-delay
lenses from current and upcoming surveys will be
realized \citep[e.g.,][]{TreuEtal13}.  We are entering an exciting era of \textit{accurate} cosmology as various methods begin to gain both the 
precision and accuracy required to rule out cosmological models and potentially
discover new physics.

%-------------------------------------------------------------------------------

\section*{Acknowledgments}

We thank Adriano Agnello, Peter Schneider, Dominique Sluse, Kenneth
Wong and the anonymous referee for useful 
comments/discussions.  We acknowledge support by NASA through \hst\ grant GO-12889.
S.H.S.~acknowledges support from the Ministry of Science and Technology in Taiwan via grant MOST-103-2112-M-001-003-MY3.
T.T.~acknowledges support from the Packard Foundation and the National Science Foundation (NSF) grant 1312000. 
R.D.B.~acknowledges support from NSF-AST-0807458.
F.C., G.M., and M.T.~acknowledge support from the Swiss National Science 
Foundation.
C.D.F.~acknowledges support from NSF-AST-0909119.
This paper is based in 
part on observations made with the NASA/ESA \hst, obtained at the Space Telescope Science Institute, which
is operated by the Association of Universities for Research in
Astronomy, Inc., under NASA contract NAS 5-26555. 
These observations
are associated with program GO-9744.

%-------------------------------------------------------------------------------
% bibliography:

%\bibliography{RXJ1131paper2}
%\bibliographystyle{apj}

%-------------------------------------------------------------------------------

\label{lastpage}
\end{document}